\documentstyle[12pt]{article}
\newcommand{\beq}{\begin{equation}}
\newcommand{\eeq}{\end{equation}}
\newcommand{\bea}{\begin{eqnarray}}
\newcommand{\eea}{\end{eqnarray}}

 \newcommand{\reset}{  \setcounter{equation}{0}  }
\newcommand{\th}{\theta}

\newcommand{\ib}{ \bar{ \imath}}
\newcommand{\jb}{ \bar{ \jmath}}
\newcommand{\etab}{ \bar{ \eta}}

\newcommand{\xh}{ \hat{ x}}

\newcommand{\PL}{{\em Phys. Lett. }}
 \newcommand{\NP}{{\em Nucl. Phys. }}
 
\newcommand{\CMP}{{\em Comm. Math. Phys.  }}
\begin{document}
\setcounter{page}{0}
\topmargin 0pt
\oddsidemargin 5mm
\renewcommand{\thefootnote}{\fnsymbol{footnote}}
\newpage
\setcounter{page}{0}
\begin{titlepage}
\begin{flushright}
Berlin Sfb288 Preprint No. 157 \\
hep-th/9503019
\end{flushright}
\vspace{0.5cm}
\begin{center}
{\Large {\bf Braid Relations in Affine Toda Field Theory} }

\vspace{1.8cm}
{\large Andreas Fring}
\footnote{e-mail address: Fring@omega.physik.fu-berlin.de}  \\
\vspace{0.5cm}
{\em  Institut f\"ur Theoretische Physik\\
Freie Universit\"at Berlin, Arnimallee 14, D-14195 Berlin, Germany}

\end{center}
\vspace{1.2cm}

\renewcommand{\thefootnote}{\arabic{footnote}}
\setcounter{footnote}{0}

\begin{abstract}
We provide explicit realizations for the operators which when exchanged give
rise to the scattering matrix. For affine Toda field theory we present two
alternative constructions, one related to a free bosonic theory and the
other formally to the quantum affine Heisenberg algebra of $U_q(\hat{Sl_2})$.
\end{abstract}
\vspace{.3cm}
\centerline{March 1995}
\end{titlepage}
\newpage
\section{Introduction}

Integrable theories in 1+1 space-time dimensions represent the hitherto best
understood
examples of relativistic quantum field theories. Whereas for the classical
theories integrable
refers to the existence of an infinite number of conserved quantities, such
that the equations
of motion are solvable, in the quantum theory one expects that the eigenvalue
spectrum of
the Hamiltonian can be determined.
\par
Conformal field theories are particularly well understood examples of this
concept. They may
be viewed in two alternative ways, either one may view the massive theories
as perturbations
of the conformally invariant theories \cite{Zamol} and thus explain the
origin of mass, or
one may regard the massless theories as limiting cases of the former. Toda
field
theories \cite{MOP} are relativistically invariant scalar field theories
which provide
elegant unifying examples allowing for both viewpoints. Many of its
characteristics reflect
properties of the underlying Lie or affine Lie algebra for the massless or
massive case already at the classical level,
respectively. Due to the remarkable renormalisation properties
many quantities survive the quantization procedure, in particular the classical
mass ratios for theories related to simply laced Lie algebras.
\par
Since the formulation of the Wightman axioms it has been attempted to set up
axiomatic
systems which are so restrictive, that they allow for a unique construction
of an
exact quantum field theory, opposed to a perturbative approach.
The bootstrap program \cite{ZZ} provides such a system and permits to determine
the scattering matrix. This approach may in principle be extended to the
situation
off-shell \cite{Karowski,Smirnov} and allows to compute form-factors and hence
correlation functions. Recently the extension of this approach to the situation
in which boundaries are present has attracted much attention.
\par
All axioms may be derived when one postulates the existence of a certain set
of operators, which implies that they already contain all the information of
the theory. It seems therefore desirable to obtain explicit representations
for these operators and illuminate their structure, which in particular with
regard to the form-factors may lead to their explicit construction.
\par
The paper is organized as follows: In section 2 we state the general
properties of the
operators and in section 3 we demonstrate how they may be employed to derive
the
axiomatic system for the scattering matrix. We illustrate this approach with
the
explicit example of affine Toda field theory for which we present in section 4
a derivation of the expression for the S-matrix as a phase and in section 5
and 6
explicit realizations of these operators. In section 7 we state our
conclusions.
\section{Zamolodchikov algebra}
\reset
About sixteen years ago it was suggested by Zamolodchikov and Zamolodchikov
\cite{ZZ}
that one may place a postulate concerning the existence of a certain set of
creation and
annihilation type operators $Z_i( \th_i)$ and $Z^{\dag}_i(\th_i)$ at the
center
of the
formulation of a massive quantum field theory in 1+1 space-time dimensions.
These operators
are
associated to the particles of the theory in a one-to-one fashion and are
characterized
by  their quantum number $i$ and their dependence on the rapidity $\th_i$.
It is
common
to parameterize the two-momenta of the particle as $p_i = m_i (\cosh \th_i,
\sinh \th_i )$
since then the branch cuts in the complex Mandelstam s-plane unfold. Any
product
of
these operators is thought to constitute an element of an associative algebra.
Assuming
the existence of a vacuum $|0 \rangle$ one may construct a Hilbert space by a
successive
action of   $Z^{\dag}_i(\th_i)$ on it
\beq
Z_1^{\dag}( \th_1)  \ldots  Z_n^{\dag}(  \th_n)  |0 \rangle \;\;  .
\eeq
Analogously one defines the dual space to this by acting with $Z_i(
\th_i)$
successively from  the right on  $ \langle  0|$. For the  in- and
out-states one can
then select a set of linearly independent vectors in the way, that one orders
the in-states
with decreasing and the out-states with increasing  real parts of the
rapidities.
Every state which is not ordered in this way is regarded as being neither an
in-
nor an out-state.
\par
Then the n-particle scattering matrix is  defined to be the unitary matrix
which
relates the
in- and the out-states
\beq
Z_n^{\dag}( \th_n)  \ldots  Z_1^{\dag}( \th_1)  |0 \rangle_{\rm out} =
\prod_{1 \leq i<j \leq n} S_{ij} (\th_{i} - \th_{j} )  Z_1^{\dag}( \th_1)
 \ldots  Z_n^{\dag}(
\th_n)  |0 \rangle_{\rm in}
\;\;  , \label{eq: defs}
\eeq
with $ {\rm Re} (\th_1) > \ldots > {\rm Re} (\th_n$).
Here we have assumed that the quantum numbers of the in- and out-state do not
differ, which implies that the S-matrix is diagonal. This is the case for the
theories we are concerned with in the following, i.e. affine Toda field
theory with real coupling constant. Furthermore
we have employed the fact
that in an
integrable theory the n-particle S-matrices  factorize into $n(n-1)/2$ two
particle-S-matrices
and it will therefore be sufficient to characterize their properties.  Thus
one
may view the
n-particle scattering as the outcome of the re-ordering of the operators
$Z^{\dag}$ or
more specifically for an integrable theory one may regard the two-particle
scattering
matrix as the result of  the braiding of two of these operators
\bea
Z_i (\th_i) Z_{j}(\th_j) &=& S_{ij}(\th_{i} - \th_{j} ) \, Z_j(\th_j)
Z_i (\th_i)
\label{eq: Zamalga}     \\
Z_i^{\dag}(\th_i) Z_j^{\dag}(\th_j) &=&S_{ij}(\th_{i} - \th_{j} )\,
Z_{j}^{\dag}(\th_j)
Z_i^{\dag}(\th_i)
 \label{eq: Zamalgb} \;\;  .
\eea
It remains to be specified what the braiding of  $Z$ with $Z^{\dag}$ will lead
to. Viewing these operators as a generalization of a Bose-Fermi algebra, one
may postulate that
\beq
Z_i(\th_i) Z_j^{\dag}(\th_j) = S_{ij}(\th_{j} - \th_{i} ) \, Z_j^{\dag}(\th_j)
Z_i(\th_i) +  \;\delta_{ij} \;
\delta({\th_i - \th_j}) \, . \label{eq: Zamalgc}
\eeq
In this case one recovers for $S \rightarrow \pm 1$ the usual bosonic
commutation and fermionic anticommutation relations, respectively. This limit
may be obtained either by letting the rapidity go to zero or the effective
coupling to the value corresponding to the free theory.
Viewing the asymptotic states as representations of the Z's, we may
interpret $Z_i^{\dag}(\th)$ and $Z_i(\th)$ as particle creation and
annihilation
operators, respectively. Besides the fact that one needs relations
connecting the Hilbert space with its
dual, the
latter equation  may be employed as well for the derivation of
form-factor axioms
\cite{Karowski,Smirnov,Smirnov1,YurovZ}, that is in particular for the
derivation
of the kinematic residue equation (refer appendix of \cite{Smirnov}).
\par
In addition to the braiding properties we require a characterization of the
singularity structure
of the Z's in order to obtain a complete description.
Annihilation processes are characterized by the operator product expansion
\beq
\lim_{\bar{\th} \rightarrow \th} Z_i( \bar{\th} + i \pi ) Z_{ \ib} ( \th ) =
  c_i
\;\;   ,\label{eq: opeann}
\eeq
with $c_i$ being some  constant. Clearly the same relations hold for
$Z_i^{\dag}(\th)$.
\par
Possible fusing processes, $ Z_i + Z_j \rightarrow Z_k$,  are incorporated via
 the following  operator product expansion
\beq
 \lim_{ \bar{\th} \rightarrow \th } ( \bar{ \th} - \th  ) Z_i \left( \bar{\th}
 + i \bar{\eta}_{ik}^{j} \right) \;
Z_j \left( \th - i \bar{\eta}_{kj}^{i} \right) = \Gamma_{ij}^k Z_k( \th )
 \;\;  . \label{eq: opebound}
\eeq
 Here
the $   \bar{ \eta}_{ij}^k =  \pi  - \eta_{ij}^k$    denote the fusing
angles for bound
states and  $\Gamma_{ij}^{k} $  the three-particle  vertex on mass-shell
defined via
\beq
\rm{Res} \;\; S_{ij}(\th) |_{i \eta_{ij}^k } =  \left( \Gamma_{ij}^k \right)^2
 \;\; . \label{eq: bound}
\eeq
\par
One may still find solutions to the preceding set of equations which will
however
lack a definite interpretation as a quantum field theory. One further
restriction
emerges from the existence of higher order poles which have to be given an
interpretation in form of a generalization of the Coleman-Thun mechanism
\cite{Cole,CM}. On the level of the scattering matrix it is in general a very
cumbersome procedure to verify whether all the required graphs exist and a
more concise argumentation is highly desirable. For the
operators of the Zamolodchikov algebra we expect for a second order poles
\beq
{\rm Res}|_{\th = \th'}\; Z_i(\th + i \eta)  Z_j(\th' - i \eta') \; = \;
\Gamma_{i \bar{k}}^l \Gamma_{jk}^m  \; Z_l(\th) Z_m(\th)
\label{eq: seboot}
\eeq
whereas a third order pole may be incorporated into
\beq
{\rm Res}|_{\th = \th'}\; Z_i(\th + i \tilde{\eta})  Z_j(\th' - i
\tilde{\eta}')
\; = \; \Gamma_{i \bar{k}}^l \Gamma_{jk}^m \Gamma_{lm}^n \; Z_n(\th) \;\; .
\label{eq: thboot}
\eeq
The angles $\eta + \eta'$ and $\tilde{\eta} + \tilde{\eta}'$ correspond to the
second and third order pole, respectively, in the scattering matrix $S_{ij}$
and have definite relations to the more fundamental fusing angles which
occur in (\ref{eq: bound}). In general one obtains for an $N^{\rm th}$ order
pole
\beq
{\rm Res}|_{\th = \th'}\; Z_i(\th + i \hat{\eta})  Z_j(\th' - i \hat{\eta}')
\; = \; \Gamma_{1} \ldots \Gamma_{N}   \prod_l  Z_l(\th)  \;\; .
\label{eq: Nboot}
\eeq
\par
Further restrictions on possible fusing processes and therefore on the
operators result when a symmetry is present in the theory. For instance
invariance of the operator product expansion under $Z(N)$-symmetry,
i.e. $Z_l(\th) \rightarrow \exp \frac{l \;2 \pi i}{N} Z_l(\th)$
\cite{KS} will put severe restrictions on the possible labels for the
Z's. In affine Toda theories these symmetries may all be identified with
the automorphisms of the Coxeter-Dynkin diagrams.
\par
The associativity of the algebra  (\ref{eq:  Zamalga}) - (\ref{eq: Zamalgc})
together with the
operator product expansion imply certain consistency conditions for the
two-particle S-matrix,
like the Yang-Baxter equations  \cite{YB} and the bootstrap equations
\cite{ZZ}, as we shall
demonstrate in the next section. It is worth noting that this ideas  extend to
the case in which we have a reflecting boundary in space simply by adding
one  more
generator to the algebra which then represents a wall or a defect \cite{FK}.
These operators are  however
not quite on the same footing, since for instance for the case of pure
reflection, they will always remain on the right in every  product.
\par
Having broken the conformal symmetry one nonetheless expects the presence of
an
operator,
say $d'$, which generates a Lorentz boost
\beq
e^{\Delta d'} Z_i(\th) e^{-\Delta d'} \; = \; Z_i(\th  + \Delta) \;\; .
\eeq
This is realized if it satisfies the relation
\beq
   [ d', Z_i(\th) ] \; = \; \frac{ d}{ d \th} Z_i(\th)  \;\;\;   .
\label{eq: boost}
\eeq
In addition, due to the integrability of the theory under consideration, we
presume that by
integrating a certain set of local densities we obtain an infinite set of
conserved, real and
commuting charges $Q_s$, which are graded by their spin $s$.  The possible
values
of this spins depend naturally on the theory under consideration, for instance
in affine Toda field theories, the values of  this spins are known to equal
the exponents
modulo the  Coxeter
number $h$ of the  underlying
Lie algebra \cite{OT} at the  classical level. The charges are expected to
act in the space of states
(\ref{eq: defs}) as well and are assumed to
diagonalise them \cite{ZZ}
\beq
Q_s \; Z_1^{\dag}( \th_1)  \ldots  Z_n^{\dag}( \th_n)  |0
\rangle_{\!\! {\rm in}
\atop{\rm out} } \; =\;
\sum_{i=1}^{n}
\omega_s^i(\th_i) \; Z_1^{\dag}( \th_1)  \ldots  Z_n^{\dag}( \th_n)  |0
\rangle_{\!\! {\rm in}
\atop{\rm out} }   \;\; .
\eeq
Carrying out a spatial reflection relates positive and negative spins and we
may derive the
important relation for the eigenvalues of $Q_s$ on the one-particle states
\beq
\omega_s^i(\th_i) \; = \; \omega_{-s}^i(-\th_i) \;\;\; .   \label{eq: spineig}
\eeq
Acting with $Q_s$ on the representation of (\ref{eq: opebound}) we
immediately obtain a relation found
originally in \cite{Zam}
\beq
\omega_s^i (i \bar{\eta}_{ik}^j ) + \omega_s^j (-i \bar{\eta}_{kj}^i ) =
\omega_s^k (  0  ) \;\;  .
\label{eq: spinc}
\eeq
On the other hand, acting on the
representation of (\ref{eq:  opeann}) one obtains
\beq
\omega_s^i(0) \; = \; (-1)^{1 + s} \omega_s^{\ib}(0) \;\; ,
\eeq
which yields the well known relation for the electrical charge with $s=0$
and the masses with $s = 1$.
This equation may be employed as a very powerful tool, i.e. knowing the
values of the spins $s$ one is able to solve these equations for the
possible values of $\bar{\eta}$ and hence construct the S-matrix via the
bootstrap principle \cite{Zamol}.
\par
Although the concept described above is relatively old, it has remained on a
rather formal
level and explicit realizations of the Zamolodchikov algebra, in particular
for Lagrangian field theories, have hitherto been
found only
for few theories \cite{Smirnov2,Kulish,CD,LU,Ch,LU2}.
\par
Evidently such an algebra will crucially  depend on the explicit nature of
the theory and
in order to proceed  with an explicit construction one has to consider a
concrete theory.
Nonetheless, without possessing any explicit realization, one is able to
make a few model independent statements.  It is instructive to see how
the previous relations lead to axiomatic systems for the on-shell quantities
of any integrable theory. This may also be extended to the situation
off-shell and
a realization of the algebra may be employed to construct explicit solutions
for instance for the form-factors in form of trace functions \cite{LU}.

\section{S-matrix axioms}
\reset

For the case at hand, an integrable quantum field theory, the on-shell
properties are
characterized by the two-particle scattering matrix for which the axioms
arise in a
straightforward manner. Parity invariance and ``unitarity" may be obtained
without having
any explicit
realization for the Z's in a fairly simple way.
Employing (\ref{eq:  Zamalga}) twice for
$Z_i(\th_i)$ $Z_j(\th_j)$  yields
\beq
S_{ij} (\th_{ij} )  S_{ji} (\th_{ji} )  \; = \; 1   \;\;   .
\eeq
This equations may be satisfied  if both parity, i.e. symmetry in $i$
and $j$ and ``unitarity"
hold
\beq
S_{ij} (\th_{ij} ) \; = \; S_{ji} (\th_{ij} ) \; = \;  S_{ij} (\th_{ji} )^{-1}
  \;\;  . \label{eq: paruni}
\eeq
By Lorentz invariance, the scattering matrix depends only on the rapidity
difference
which we denote from now on by $\th_{ij} := \th_i - \th_j$.
Similarly straightforward is to derive the crossing relation, which relates
the
scattering of particle $i$ and $j$ to the scattering of $i$ with $\jb$.
Considering
\beq
 Z_j( \th_j + i \pi ) Z_{\jb} ( \th_j ) Z_i( \th_i )
\eeq
we may by associativity either annihilate the $Z_j$ and $Z_{\jb}$ first by
relation (\ref{eq: opeann})
or commute the $Z_i( \th_i) $ to the left and then contract.
Equating the results implies the crossing relation for the S-matrix
\beq
S_{i \jb} ( \th_{ij} + i \pi)  S_{ij}( \th_{ij} ) \; =\; 1 \;\; .
\eeq
Finally one may derive the bootstrap equation  for the scattering matrix
\beq
S_{li} \left(\th + i \bar{\eta}_{ik}^j  \right) \; S_{lj} \left(\th - i
\bar{\eta}_{kj}^i \right) \;
= \;  S_{lk} \left(\th \right) \;  \;\;  .
\eeq
By considering the product
\beq
Z_i  \left( \bar{\th} + i \bar{\eta}_{ik}^j  \right)  \;
Z_j\left( \tilde{\th} -
i  \bar{\eta}_{kj}^i \right) \;
 Z_l (\th') \;\;  , \label{eq: stt}
\eeq
we may either commute $Z_l$ with $Z_i$, $Z_j$ and then compute the residue
for $ \bar{\th} \rightarrow
\tilde{ \th} $ or  alternatively contract first $Z_i$ and  $Z_j$ and
subsequently commute
 $Z_{k} $ with  $Z_l$. Then  upon denoting $ \th =  \tilde{ \th }   - \th' $
and equating the results
leads directly to the bootstrap equation for the S-matrix.
\par
When the residue (\ref{eq: bound}) is positive, having assumed a bosonic
theory, the poles $i \eta_{ij}^k$ are related to the
bound state $Z_k$ in the direct channel of the process $Z_i + Z_j$,
satisfying the usual
relation for the masses of the fusing particles \cite{ZZ}
\beq
m_k^2 = m_i^2 + m_j^2 + 2 m_i m_j \cos \eta_{ij}^k \;\;  .
\eeq
For this equation to hold it is crucial that all masses renormalise uniformly,
 i.e.
the classical mass ratios pertain. This is a somewhat unappealing feature which
still plagues the bootstrap approach, that one still has to resort to
perturbative
arguments in order to judge whether a solution to the set of axioms is suitable
or not. One further restriction emerges for theories with higher order poles,
like for instance affine Toda field theories in which we may
have poles up to order 12. Then the Coleman-Thun mechanism \cite{Cole}
may be generalized to the extend that all odd order poles with
positive residues are permitted to participate in the bootstrap \cite{CM,BCDS}.
The bootstrap equations corresponding to this poles may be derived from
(\ref{eq: stt}), after suitably re-interpreting the fusing angles,
together with
(\ref{eq: seboot}) - (\ref{eq: Nboot}) in the same fashion. Hence one has
excluded
all redundant poles and zeros. Further ambiguities, the so-called CCD-factors
\cite{CDD},  may be excluded by the additional assumption that
$-i \ln S( \th) $ is damped for $\th \rightarrow \infty$.
\par
In a similar fashion one may employ the associativity of the algebra and
derive
Yang-Baxter equations, which however, due to the diagonality of the two
particle S-matrices
we are concerned with, only contain trivial information and may therefore
be ignored in the present context.


\section{The S-matrix as a phase}
\reset

For our purpose and in various other contexts,  like for instance when
computing
finite size effects via the
Thermodynamic Bethe Ansatz,  one  requires the S-matrix to be represented in
 form of a phase.
In this section we therefore present an explicit derivation of it.
Adopting the notation of \cite{FLO,FO} the S-matrix of affine Toda field
theories
related to simply laced Lie algebras
\cite{AFZ,KS,Zamol,FZ,DDV,FKM,CM,BCDS,PD,FO,Musrep}
may be  cast into the very compact form
\beq
S_{ij}(\th) \; = \; \prod_{q=1}^{h} \left\{ 2q - \frac{ c(i) + c(j)}{2}
\right\}_{\th}^{-  \frac{1}{2} \lambda_i \cdot \sigma^q \gamma_j }  \; = \;
 \frac{ \chi_{ij}(\th ) }
{ \chi_{ij}(-\th ) }\;\;\; ,
\label{eq: smat}
\eeq
with
\bea
\chi_{ij}(\th )  & := &  \prod_{q=1}^h \left[ 2q + 1 + \frac{ c(i) - c(j)
}{  2} \right]_{ \th }^
{ \sigma^q \lambda_i \cdot \lambda_j  }  \\
& = &  \prod_{q=1}^h \left(   \frac{ \left( 1 - \omega^{-q} e^{- \th -
\frac{ i \pi}{h} \left( 2+ \frac{c(i)
-c(j)}{2}   \right) } \right)  \left( 1 - \omega^{-q} e^{- \th -
\frac{ i \pi}{h} \left( \frac{c(i)  -c(j)}{2}
\right) } \right)  } { \left( 1 - \omega^{-q} e^{- \th -  \frac{ i \pi}{h}
\left( 2 -B + \frac{c(i)
-c(j)}{2}  \right)  }  \right) \left( 1 - \omega^{-q} e^{- \th -
\frac{ i \pi}{h} \left( B +  \frac{c(i)  -c(j)}{2}
  \right) } \right) }  \right)^{ \sigma^q \lambda_i \cdot \lambda_j  }
\nonumber \;\;   .
  \eea
 The  $\{ \}_{\th}$ are building blocks consisting  out of sinh-functions,
i.e. $\{ x\}_{\th} = [ x ]_{\th}/ [x]_{-\th}, [x]_{\th} = < x+1>_{\th}
<x-1>_{\th} / < x+1-B>_{\th} <x-1+B>_{\th} $ and $< x>_{\th} = \sinh\frac{1}{2}
\left( \th + \frac{i \pi x}{h}\right)$. $B(\beta) =
\frac{2 \beta^2}{4 \pi + \beta^2 } $ is  the effective
coupling constant  which takes values between 0 and 2 when $\beta$
is taken to be purely real. $h$ denotes the Coxeter number and $\omega$ is
the $h^{\rm th} $ root of
unity.   $\sigma$ denotes
the particular Coxeter element in the conventions of \cite{FO}, $c(i)$ the
colour values related
to the bicolouration of the Dynkin diagram, $\lambda_i$ a fundamental weight
and $\gamma_i$
is $c(i)$ times a simple root.
\par
In \cite{FJKO} it was noted that the function  $ X_{ij} (z,\xi)  =
 \prod\limits_{q=1}^h \left( 1 -
 e^{ - \frac{2 \pi i }{h}  q}  \left(  \frac{ \xi}{z}  \right)
\right)^{  \sigma^q \gamma_i \cdot \gamma_j }$, which arose from the normal
ordering of two
vertex operators associated with classical solitons, exhibits close
similarities to the
two-particle scattering amplitude. The precise relation will become more
apparent in this
section.  In analogy to the expressions of section 5.2 in \cite{OTU}   we
obtain for
 $ | \xi | < | z |  $ the identity
\beq
\exp\left(  - \sum_{ n \in {\bf e} } \frac{ \left( \lambda_i \cdot q^*(n)
\right)
 \left( \lambda_j \cdot q(n) \right) } {n}    \left(   \frac{ \xi}{z}
 \right)^n
\right) \; = \;
\prod_{q=1}^h \left( 1 - e^{ - \frac{2 \pi i }{h}  q}  \left(  \frac{ \xi}{z}
\right)
\right)^{  \sigma^q \lambda_i \cdot \lambda_j } \;\;   .
\eeq
Here the $q(n)$ form an orthogonal and complete set of eigenvalues of the
Coxeter element
with eigenvalue $ \omega^{n} $ and  ${\bf e}$ denotes a set of
positive integers,
the exponents of the Lie algebra modulo $h$. Employing this relations,  a few
manipulations
lead to
\beq
 \ln \chi_{ij}(\th_{ij} )  =   4 \sum_{ n \in {\bf e} } \frac{ \left(
\lambda_i
\cdot q^*(n) \right)
 \left( \lambda_j \cdot q(n) \right) } {n} e^{-n   \left(  \th_{ij} -
\frac{i \pi}{h}
\left( \frac{c(i)  -c(j) }{2} +1
\right) \right) }  \sin\frac{n \pi}{2 h} B \sin (2-B)\frac{n \pi}{2 h}  .
\eeq
Keeping in mind that we used  $ {\rm Re}( \th_i )> {\rm Re} (\th_j) $ we
cannot
simply interchange
the two rapidities in order to derive the expression involving
$\chi_{ij}(\th_{ji} ) $.
Careful analysis then yields for the phase of the S-matrix
defined as $ \delta_{ij}(\th) =   -i \ln S_{ij}(\th)$
\bea
\delta_{ij}(\th )   & = &  \varepsilon_{\th} \sum_{ n \in {\bf e} }  f(n,B)
\frac{ \left(
\lambda_i \cdot q^*(n) \right)
 \left( \lambda_j \cdot q(n) \right) } {n} e^{-n    \left(\varepsilon_{\th}
\th
- \frac{i \pi}{h}
\left( \frac{c(i)  -c(j) }{2}  \right)  \right) } \label{eq: lnSw}    \\
& = & \varepsilon_{\th}  \sum_{ n \in {\bf e} }  \frac{ \xh_i(n)\xh_j(n)}{n}
f(n,B)
  e^{-n \varepsilon_{\th} \th }    \label{eq: lnSx}
\eea
with
\beq
f(n,B) := \sin\frac{n \pi}{2 h} B  \; \sin (2-B)\frac{n \pi}{2 h}
\label{eq: Vorfaktor}
\eeq
and $\varepsilon_{\th} = \pm 1$ for ${\rm Re} \th > 0 $ and ${\rm Re}
\th < 0 $,
respectively.
We have evaluated the inner product of the
fundamental weights and the eigenvectors of the Coxeter element
\beq
 \sigma^l\lambda_i \cdot q(n) = x_i(n) \exp\left( - \frac{i \pi}{h} n \left(
\frac{c(i) +1}{2}  + 2l
\right)   \right)  \; \; . \label{eq: weightq}
\eeq
The $x_i(n) $ denote the left eigenvectors of the Cartan matrix and
 $\xh_i(n) = \sqrt{ 2/ \sin(n\pi /h)} x_i(n) $.
This expression may be derived from  the relation between the fundamental
weights
and roots $ \gamma_i = ( \sigma_-   - \sigma_+)  \lambda_i$   and the inner
product of   $  \sigma^l\gamma_i \cdot q(n) $ \cite{FO}. Similar expressions
for S have been quoted in \cite{KM,Nie}.
\par
In order to convince ourselves that no mistakes have been made in the
derivation
we can check for consistency whether (\ref{eq: lnSw}) and (\ref{eq: lnSx})
still
satisfy the S-matrix axioms.   First one may verify the crucial relation
\footnote{
In general we have that $ \lim\limits_{\th \rightarrow 0} \{ x \}_{\th} =1 $,
except for $x=1$ in which case we obtain $-1$. Due to the presence of the
block $\{ 1 \}_{\th}$ only in $S_{ii}$ one obtains $ S_{ij}(0) =
(-1)^{\delta_{ij}}$.  }
\beq
\lim_{\th \rightarrow 0} \delta_{ij}(\th) \; =\; - \delta_{ij} \; \pi \; .
\eeq
We do not have an analytic proof of this relation, but it may be verified
easily numerically. This relation confirms in particular that  we have
chosen the correct normalization.
The parity  invariance, i.e. the
symmetry in i and j, and the unitarity relations   (\ref{eq: paruni})
follow trivially from  (\ref{eq: lnSx}).
Not quite so  obvious is the crossing relation. Relating the fundamental
weights
of the anti-particles to the ones of  the particles in a
similar way as the representatives of the orbits $ \Omega_i$ and
$\Omega_{\ib}$
\cite{FLO}, we
obtain
\beq
\lambda_{\ib} = - \sigma^{ -\frac{h}{2} + \frac{ c( \ib) - c(i) }{4} }
\lambda_i
\eeq
and hence the crossing relation
\bea
S_{i \jb}( \th_{ij}) & = &\exp\left( - \sum_{ n \in {\bf e} }  f(n,B)  \frac{
\left(  \lambda_i \cdot q^*(n) \right)
 \left( \lambda_j \cdot  \sigma^{ \frac{h}{2} + \frac{ c(j) - c(\jb) }{4} }
q(n) \right) } {n} e^{-n   \left(  \th_{ij}
- \frac{i \pi}{h}  \left( \frac{c(i)  -c(j) }{2}  \right)  \right) }  \right)
   \nonumber \\
 &= &S_{i j }( i \pi -\th_{ij})  \;\; .
\eea
Alternatively, we could have used $ x_{\ib}(n)  = - x_i(n) e^{n i \pi} $
in (\ref{eq:  lnSx}).
Finally we would like to verify the bootstrap equation, which after the usual
identification of the fusing angles is most suitable in the form
\beq
 \prod_{t= i,j,k}  S_{lt} \left(\th + i \eta(t) \right)    \; =\; 1  \; .
\label{eq: boot}
\eeq
Here the fusing angles are given by $\eta(i) = - \frac{\pi}{h} \left( 2 \xi(i)
+ \frac{ 1 - c(i) }{2}
\right) $ and the integers $ \xi(t) $  comprise two independent conjugacy
classes
of integers which
provide  solutions  to the fusing
rule \cite{PD,FLO,FO} $ \sum\limits_{t=i,j,k} \sigma^{- \xi(t) } \lambda_t =
0$.
Then considering
$$
\lambda_t \cdot q(n) e^{ - n \th - i n \eta(t) } e^{ n \frac{i \pi}{h} \left(
\frac{ c(t) - c(l)}{2} \right)} =
\sigma^{- \xi(t) } \lambda_t  \cdot q(n) e^{- n \frac{i \pi}{h} \left( \frac{
1+
c(l)}{2} \right)}
$$
yields, by summing over $i,j,k$ and by employing the fusing rule, precisely
the
bootstrap
equation (\ref{eq: boot}). So indeed all the S-matrix axioms are satisfied. We
shall be
content here with this check and do not report on the pole structure of
 (\ref{eq:
lnSw}) or (\ref{eq: lnSx}).
\par
Using the fact that $f(n,B) = - \frac{ n \beta^2}{4 h} \sin \frac{ n \pi}{h}
+ {\cal{O}}( \beta^4) $ one
obtains the precise relation between the function $X_{ij}(\th)$
\cite{FJKO} which played an
important role in the study of the classical soliton solutions and the phase
of the scattering matrix
\beq
\delta_{ij}(\th) \; = \; - \frac{ \beta^2}{2 h} \frac{d}{d\th} \ln X_{ij}(\th)
 \;\; .
\eeq
This is the reversed situation as in the soliton case, in which the phase shift
may be obtained by integration of the classical time delay \cite{JackiwWoo}.


\section{Bosonic Representation}
\reset
As already mentioned the algebra (\ref{eq: Zamalga}) - (\ref{eq: Zamalgc})
may be regarded as a generalization of a Bose-Fermi algebra. It is known
since some time that it is possible to find operators inside the bosonic Fock
space which obey fermionic commutation relations and vice versa there exist
operators of bosonic nature which may be expressed in terms of fermionic
operators \cite{BFOP}. This observation, among others, led to the
notion that in two dimensional space-time bosons and fermions are regarded as
equivalent. In \cite{KT} it was shown that this construction may be
generalized to arbitrary spins. It is interesting to note that one may
further extend this in such a way that one obtains
an explicit representation
of the Zamolodchikov algebra in terms of free bosonic fields $a_i(\th)$.
This may be achieved by replacing the spin by the rapidity
dependent phase of the S-matrix and therefore turning the conventional
expression into a convolution. So defining now the operators
\bea
Z_i(\th) &:=&  a_i(\th) \;  \exp\left( i
\int_{\th }^{\infty} d\th'
\delta_{il} (\th - \th') \; n_l(\th') \right)  \\
Z_i^{\dag}(\th) &:=&  a_i^{\dag}(\th) \;
\exp\left(- i \int_{\th }^{\infty}
d\th' \delta_{il} (\th - \th') \; n_l(\th') \right)
\eea
we have obtained a solution of the algebra
(\ref{eq: Zamalga}) - (\ref{eq: Zamalgc}) in terms of free bosonic operators,
which are assumed to satisfy as usual
\bea
\; [ a_i(\th) , a_j^{\dag}(\th') ] &=& \delta_{ij} \; \delta(\th - \th') \\
  \left[ a_i(\th) , a_j(\th') \right] &=& \left[ a_i^{\dag}(\th) ,
a_j^{\dag}(\th') \right] \;=\; 0
\;\;\; .
\eea
One may verify this as usual by employing the Baker-Campbell-Hausdorff
formula. Here we denote with $n_i(\th) = a_i^{\dag}(\th) a_i(\th)$ the number
density
operator.
The scattering matrix should then be of the form
\beq
\ln S_{ij} (\th) \;=\; i \varepsilon_{\th} \delta_{ij}( \varepsilon_{\th} \th )
\;\; .
\eeq
For vanishing effective coupling one expects that,
$\delta \rightarrow \pi n$, such that the algebra reproduces the usual
bosonic or fermionic algebra. This is indeed the case when $\delta_{ij}(\th)$
is taken to be the phase of the S-matrix of affine Toda field theory, as may be
verifies easily with the expressions of the previous section. This solution
is completely model independent and may be employed for all diagonal theories.
On the possible extension to the non-diagonal case we report elsewhere.

\section{Vertex operator construction }
\reset
When relaxing relation (\ref{eq: Zamalgc}) and solely demanding that the
braiding
of two Z's will pick up the scattering matrix one may construct conceptually
entirely different representations. In fact this relation was only employed
in order to derive the kinematic residue equation for the form factors.

\subsection{The exchange relations for the Z's}
We shall now try to find explicit representations for the operators Z which
obey
the algebra
(\ref{eq: Zamalga}) for affine Toda field theories. Such a construction has
originally been suggested by Corrigan and Dorey
\cite{CD}.  Here we shall  propose an alternative construction.
Due to the close resemblance of the factor $X_{ij}(\th)$ and the phase of the
scattering matrix, which was demonstrated in section 4 it is suggestive
to  consider  the vertex operator construction formulated in
\cite{OTU} for the
classical solitons in affine Toda field theory. With this motivation in mind
we may take the function $F$
introduced in  \cite{OTU} as a prototype representative of the Zamolodchikov
algebra
\beq
F_i( \xi) = \exp \left( \sum_{n \in {\bf e}}  \frac{ \gamma_i \cdot q(n) }{ n}
 \xi^n
E_{-n}
\right)
  \; \exp \left( - \sum_{n \in {\bf e} }  \frac{ \gamma_i \cdot q^{*}(n) }{ n}
 \xi^{-n}
E_{n} \right)  \;\;  .
\eeq
Here $ \gamma_i$ denotes  again a representative of the orbit $ \Omega_i$
generated by
the Coxeter  element $\sigma$ by acting on the root space, $q(n)$ an
eigenvector
of this
element as introduced  in  section  4 and the $E_n$ denote
the generators of a principal Heisenberg subalgebra of level one
\beq
 [ E_n, E_m ] \; = \; n\;\; \delta_{n+m,0} \;\;\;   . \label{eq:
Heisenberg}
\eeq
In \cite{OTU} it was demonstrated that the operator product expansion of
two F's
yields
\beq
F_i(z) F_j( \xi)  = X_{ij}(z, \xi)   \; :    F_i(z)  F_j(
\xi):
\eeq
with $: \ldots :$ denoting the normal ordering with respect to the principal
Heisenberg subalgebra and
\beq
X_{ij}(z, \xi)  = \prod_{q=1}^h \left( 1 - e^{ - \frac{2 \pi i q}{h}
}
\frac{ \xi}{z}
\right)^{  \sigma^q \gamma_i \cdot \gamma_j  }  \;\;\;   . \label{eq: OPEf}
\eeq
In order to compare this expression with the one known for the  S-matrix, it
is useful
to  convert this expression into one involving sinh-functions. Choosing
$z = \exp(\th_i) $, $
\xi = \exp(\th_j) $,
employing the relation
\beq
\sigma^q \gamma_i \cdot \gamma_j \; = \; 2 \sigma^{ q +
\frac{ c(j) -c(i)}{2} }
\lambda_i \cdot
\lambda_j - \sigma^{ q + \frac{ c(j) -c(i)}{2} - 1 } \lambda_i \cdot
 \lambda_j
- \sigma^{ q + \frac{ c(j) -c(i)}{2} + 1 } \lambda_i \cdot   \lambda_j
\eeq
and shifting the dummy variable  $q$ in (\ref{eq: OPEf}) appropriately
one obtains
\beq
X_{ij} (\th) = \prod_{q=1}^h \left( \frac{ < 2q + c(i) -c(j)
>^2_{\th} }
{ < 2q + c(i) -c(j)  +2 >_{\th}  < 2q + c(i) -c(j)  -2 >_{\th}  }
\right)^
{  \sigma^q \lambda_i \cdot \lambda_j  }   \; \;   .\label{eq: xis}
\eeq
This formula already resembles
very closely the expressions for part of the S-matrix (\ref{eq: smat}) and
  by
making the assumption,
which appears now to be quite natural, that the Z's are a kind of quantum
deformation of the
F's we are forced to split up the blocks (\ref{eq: xis}) and manipulate them
individually, i.e. essentially we have to replace a root by a weight.
Substituting now the inner product $\gamma_i \cdot q(n)$ by the
normalised left
eigenvector of the Cartan matrix $\xh_i(n)$, we obtain  the
$ 2 \pi i$-periodic operator
\beq
Z_i( \th ) = \exp \left( \sum_{n \in {\bf e}}  \frac{ \xh_i(n) }{ n} e^{n
\th  }
\hat{E}_{-n}
\right)
  \; \exp \left( - \sum_{n \in {\bf e} }  \frac{ \xh_i(n) }{ n} e^{ - n \th  }
\hat{E}_{n} \right)  \;\;  .
\eeq
Assuming now that the $\hat{E}$'s obey a deformed version of the Heisenberg
subalgebra
\beq
 [ \hat{E}_n, \hat{E}_m] \; =  i n \; f(n,B) \;   \delta_{n+m,0} \;\;\;   ,
\label{eq: defHeisenberg}
\eeq
we may carry out the operator product expansion and obtain together with
(\ref{eq: lnSx}) and (\ref{eq: weightq})
\beq
Z_i( \th_i)  Z_j( \th_j) =     \chi_{ij}(\th_{ij} )   \; : Z_i( \th_i)
Z_j(  \th_j)  : \;\;  .
\eeq
Then employing the OPE in  opposite order
and canceling the normal ordered terms gives rise to the equation (\ref{eq:
Zamalga}) and we
have found indeed a representative for the Z's.
\par
The commutation relations may be  related directly to the quantum Heisenberg
algebra
\beq
\left[ \alpha_{in} , \alpha_{jm} \right] \; = \; \delta_{n+m,0}\; \frac{1}{n}
\; \left( q^{n K_{ij} } - q^{-n K_{ij}} \right)\; \left( q^{nc} - q^{-nc}
 \right)
\eeq
which was introduced by Drinfel'd \cite{DFJ} in the context of a mode
expansion for the generators of the quantum affine algebras. Here $q$
denotes the deformation parameter, $K$ the Cartan matrix of the affine Lie
algebra and $c$ the level. Considering now $U_q( \hat{ Sl_2} )$ and
identifying $ q = e^{ \frac{ i \pi B}{4h}} $, $\hat{E}_n = n \alpha_n
\sqrt{ i} $ the two algebras become formally isomorphic at the level
\beq
c \; =\; \frac{4}{B} - 2 \;\; .
\eeq
For the conformally invariant theory the phenomenon of a coupling constant
dependent central charge is known for the Virasoro algebra.  The level one
module is obtained at the self-dual point of the theory, whilst for the free
theory the level tends to zero or infinity.
The algebra exhibits explicitly
 the strong-weak duality, i.e. invariance  under   $ B  \rightarrow 2-B$,
which is
present in conformal Toda theories  \cite{Mansfield}
and is known to survive the breaking of the conformal symmetry  \cite{AFZ,KS}.
In the  limit to the free theory $ B \rightarrow 0$ or   $ B \rightarrow 2$,
the algebra
(\ref{eq: defHeisenberg}) becomes Abelian, and the Z's become representatives
of the
usual bosonic algebra.
\par
In comparison with
\cite{CD} we have overcome the  feature that the
minimal theory
and the coupling constant dependent part do not interfere with each other.
This is
achieved by  eliminating four
of the Heisenberg subalgebras required in the construction of \cite{CD}
and replacing it by one, namely (\ref{eq: defHeisenberg}).
We also do not require any "delocalisation" in the rapidities. In \cite{Nie}
a construction which employs one Heisenberg subalgebra has been proposed. Here
the effective coupling has been moved into the operators themselves, which has
a consequence that one does not obtain the bosonic algebra for the free theory.

\subsection{Operator product expansions}
 Considering now the product of  $
Z_{i} ( \th_i )
 Z_{\ib} ( \th_i  + i \pi)$ will not lead to any nontrivial information since
the S-matrix which
is picked up  by the braiding gives $ S_{i \ib} ( i \pi ) = S_{i i} ( 0 ) = -1
$. So we have to consider
explicitly the normal ordering which yields
\beq
     : \;Z_i(\th)    Z_{\ib} ( \th + i \pi ) \; :  \; =   \; 1 \;\; .
\eeq
We then obtain
\beq
Z_i(\th)    Z_{\ib} ( \th + i \pi) \; = \; \exp \left( i \sum_{n \in {\bf e} }
\frac{\xh_i^2(n) }{n} f(n,B)  \right) = c_i = i \;\; ,
\eeq
and hence confirming relation (\ref{eq: opeann}). Once again we do not have an
analytic proof for the convergence of this series, but is may be verified
numerically.
\par
We now like to verify the operator product expansion which leads to a three
particle fusing process. Considering the product
\beq
Z_i( \th + i \bar{\eta}_{ik}^j ) Z_j( \th' - i \bar{\eta}_{kj}^i )
\eeq
we obtain for $\th' \rightarrow \th$
\bea
& &  \!\!\!\!\!\!\!\!\!\!
 \exp \left( \sum_{n \in {\bf e} } \left( \xh_i(n) e^{ i n
\etab_{ik}^j } + \xh_j(n) e^{ - i n \etab_{kj}^i } \right) \;
\frac{ e^{n \th   }}{n} \hat{E}_{-n} \right) \nonumber \\
& & \!\!\!\!\!\!\!\!\!\!
 \exp \left(- \sum_{n \in {\bf e} } \left( \xh_i(n) e^{ i n
\etab_{ik}^j } + \xh_j(n) e^{ - i n \etab_{kj}^i } \right) \; \frac{
e^{ - n  \th }}{n} \hat{E}_{-n}
\right) \; \exp \left( i \sum_{n \in {\bf e} }  \! \frac{ \xh_i(n) \xh_j(n)}{n}
f(n,B) e^{-in \etab_{ij}^k } \right).  \nonumber
\eea
Upon identifying $ \etab_{ik}^j = \eta(i) - \eta(k) = - \frac{ \pi \xi(n)}{h}
\left( 2 \left( \xi(i) - \xi(j) + \frac{ c(j) - c(i) }{ 2} \right) \right)$,
noting that the second solution of the fusing rule corresponds to
$ -\eta(i) + \eta(k)$
together with  the identity
\beq
 x_k(n) = x_i(n) e^{i n( \eta(i) - \eta(k) )}
+ x_j(n) e^{i n ( \eta(j) - \eta(k) )} \label{eq: proj}
\eeq
yields $Z_k(\th)$ for the first two factors and $\Gamma_{ij}^k$ for the
latter.
Hence we confirm (\ref{eq: opebound})
and (\ref{eq: bound}). One may also proceed further and check
the operator product expansions corresponding to the higher order poles.

\subsection{Lorentz boost}
This far we have not specified the operator which generates the Lorentz boost.
 For this purpose
we define now an operator in analogy to the Sugawara-Sommerfield construction
\cite{Sug,Somm},
but now  involving only generators of the quantum affine Heisenberg algebra
\beq
d' :=   - \sum_{ n \in {\bf e} } i \;   f(n,B)^{-1}   \;
  \hat{E}_{-n} \hat{E}_{n}   \;\; .
\label{eq: Sugawara}
\eeq
Here  $f(n,B)$  denotes the function introduced in (\ref{eq: Vorfaktor}) .
Then together with
\beq
  [  \hat{E}_{ n},  Z_i( \th)   ]   \; =  \;  i  x_i(n)   f(n,B) e^{n \th}
Z_i( \th)
\eeq
it is easy to verify that $- d'$ counts the principal grade of $\hat{E}_l$
\beq
  [ d', \hat{E}_l ] \; = \; -l \hat{E}_l  \;\;  .
\eeq
 A little bit of algebra then shows that the commutator of $d'$ with
$Z_i ( \th_i)$
is the same as the derivative of it with respect to the rapidity and
hence $d'$ indeed satisfies
(\ref{eq: boost}) and may be viewed as the generator of the Lorentz boost.
\par

\subsection{Conserved Charges}
An explicit construction for the conserved charges in affine Toda field theory
has unfortunately not yet been carried out for the quantum case.
However,
guided by the fact that classically  the conserved quantities are known to be
graded by
the exponents of the Lie algebra \cite{OT} modulo $h$, it seems very
suggestive to  conjecture that they
may be realized as
\beq
Q_n \; \sim \; \hat{E}_n  \;\; .
\eeq
Notice that this is real when choosing the convention $ (\hat{E}_n)^{\dag}
= \hat{E}_n $
and a real constant of proportionality.  This
choice is possible without upsetting  (\ref{eq: defHeisenberg}).  Furthermore,
all this charges
will commute with each other for positive $n$. We then compute
\beq
 [ Q_{ n} , Z_i( \th) ]  \; \sim \;   e^{n \th}
 Z_i (\th) \;\; .
\eeq
When the constant of proportionality is even in $n$ we observe
that (\ref{eq: spineig})  holds.
Notice further that in the free theory, i.e.  $B=0$ or $B=2$, the
vertex operators
commute with the charges. Carrying out a further consistency check and
acting with
$Q_n$ on the
state
\beq
\prod_{t = i,j,k} Z_t (\th + i \eta(t) )  |0 \rangle
\eeq
implies
\beq
\sum_{t = i,j,k}    [ Q_n   , Z_t (\th + i \eta(t) ) ] \; = \; 0
\eeq
and therefore
\beq
\sum_{t = i,j,k}   x_i(n)   e^{i n \eta(t) } = 0   \;\;  .
\eeq
This equation corresponds to equation (\ref{eq: proj}), which resulted as a
projection of the
fusing rule into the velocity plane \cite{FO}.
It is interesting to note that this
equation results as well
as an ambiguity of the solution for the form factor axioms  and
corresponds to the
consistency equation derived by Zamolodchikov \cite{Zam}  to decide whether
operators of
certain spins may  be present in a particular theory.
\par
Certainly the expressions for the charges need to be put on a more solid ground
and a proper derivation of them is still required. Nonetheless, it is quite
intriguing
that the expression has aready  the expected properties.

\section{Conclusions}
We have demonstrated that it is possible to find explicit realisations
for the operators which when exchanged will give rise to the scattering
matrix. We provided two alternative constructions. One realisation may
be obtained by extending the construction which relates particles of
different spins and statistic in two dimensions. Starting from the classical
soliton operators we construct a second realisation in terms of the generators
of the quantum Heisenberg algebra related to $U_q(\hat{Sl_2})$. It was shown
that this operators contain all the information of the scattering matrix and
therefore permit to regard them as the central objects, rather than S.
\par
The operators may be employed to extract information for the off-shell
properties of the theory, like form-factors and ultimately correlation
functions. It remains as very interesting issue to clarify how the operators
may be transformed into real space and possibly utilise them in order
to obtain the braid relations for the Toda fields themselves.
A further interesting problem, on which we report elsewhere,
is provided by the opportunity to extend this construction to the
non-diagonal situation. This may solve the still outstanding problem for the
form of the scattering matrix for affine Toda field theories
related to purely complex $\beta$ by constructing
first the operators instead of the S-matrix.
\par
\noindent
{\bf Acknowledgement}: I'm grateful to M. Karowski, D. Olive and B. Schroer
for useful comments and to the Deutsche Forschungsgemeinschaft (Sfb 288)
for support.


\begin{thebibliography}{99}

\bibitem{Zamol} A.B. Zamolodchikov, {\em Int. J. Mod. Phys.} {\bf A3} (1988)
743.
\bibitem{MOP} A.V. Mikhailov, M.A. Olshanetsky and A.M. Perelomov,
{\em Comm. Math. Phys.} {\bf 79} (1981), 473; G. Wilson, {\em Ergod. Th.
Dyn. Syst.} {\bf 1} (1981) 361; D.I. Olive and  N. Turok, {\em Nucl. Phys.}
{\bf B257} [FS14] (1985) 277.
\bibitem{ZZ} A.B. Zamolodchikov and Al.B. Zamolodchikov, {\em Ann.Phys.}
{\bf 120} (1979)  253.
\bibitem{Karowski} M. Karowski and P. Weisz, {\em Nucl. Phys.} {\bf B139}
(1978) 445; B. Berg, M. Karowski and P. Weisz, {\em Phys. Rev.}
{\bf D19} (1979)  2477;  M. Karowski, {\em Phys. Rep.} {\bf 49} (1979) 229.
\bibitem{Smirnov} F.A. Smirnov {\em Adv. Series in Math. Phys.} {\bf 14}, World
 Scientific 1992.
\bibitem{Smirnov1} F.A. Smirnov, {\em J. Phys.} {\bf A17} (1984) L873;
F.A. Smirnov, {\em J. Phys.} {\bf A19} (1984) L575; A.N. Kirillov
and F.A. Smirnov, {\em Phys. Lett.} {\bf B198} (1987) 506;
A.N. Kirillov and F.A. Smirnov, {\em Int. J. Mod. Phys.} {\bf A3} (1988) 731.
\bibitem{YurovZ} V.P. Yurov and Al. B. Zamolodchikov,
{\em Int. J. Mod. Phys.} {\bf A6} (1991) 3419.
\bibitem{Cole} S. Coleman and H. Thun, \CMP {\bf 61} (1978) 31.
\bibitem{CM} P. Christe and G. Mussardo, {\em Nucl.Phys. B} {\bf B330} (1990)
465;
{\em Int. J. Mod. Phys.} {\bf A5} (1990) 1025.
\bibitem{KS} R. K\"oberle and J.A. Swieca, {\em Phys. Lett.} {\bf 86B} (1979)
209.
\bibitem{YB}  C.N. Yang, {\em Phys. Rev. Lett.} {\bf 19} (1967) 1312;
              R.J. Baxter, {\em Exactly Solved Models
              in Statistical Mechanics} (Academic Press, London, 1982).
\bibitem{FK} A. Fring and R. K\"oberle, {\em Nucl. Phys.} {\bf B421} (1994)
159;  {\em Nucl. Phys.} {\bf B419}
(1994) 647.
\bibitem{OT} D.I. Olive and  N. Turok, {\em Nucl. Phys.} {\bf B215} [FS7]
 (1983) 470.
\bibitem{Zam} A.B. Zamolodchikov, in {\em Advanced Studies in Pure
Mathematics}
{\bf 19} (1989), 641;
{\it Int. J. Mod. Phys.}{\bf A3} (1988), 743.
\bibitem{Smirnov2} F.A. Smirnov, {\em Nucl. Phys.} {\bf B337} (1989), 156;
{\em Int. J. Mod. Phys.} {\bf A4} (1989) 4213.
\bibitem{Kulish} P.P. Kulish, {\em Zap. Nauchn. Semin. LOMI} {\bf 109} (1981)
83.
\bibitem{CD} E. Corrigan and P.E. Dorey, {\em Phys. Lett.} {\bf B273} (1991)
237.
\bibitem{LU} S. Lukyanov, {\em ``Free Field Representation for Massive
Integrable Models''}, Rutgers-preprint RU-93-30 /hep-th/9307196.
\bibitem{Ch} I.V. Cherednik,{\em  Teor. i Mat. Fiz.} {\bf 43} (1980) 117.
\bibitem{LU2} S. Lukyanov and Y. Pugai, {\em ``Bosonization of ZF Algebras:
Direction Towards Virasoro Algebra''}, Rutgers-preprint RU-94-41
/hep-th/9412128.
\bibitem{BCDS}H. W. Braden, E. Corrigan, P. E. Dorey, R. Sasaki,
{\em Nucl. Phys.} {\bf B338} (1990)  689.
\bibitem{CDD} L. Castillejo, R.H. Dalitz and F.J. Dyson, {\em Phys. Rev.}
{\bf 101} (1956), 453; P.Mitra \PL {\bf 72B} 62.
\bibitem{FLO} A. Fring, H.C. Liao and D.I. Olive, {\em Phys. Lett.} {\bf B266}
  (1991) 82.
\bibitem{FO} A. Fring and D.I. Olive, {\em Nucl. Phys.} {\bf B379} (1992) 429.
\bibitem{AFZ} A.E. Arinshtein, V.A. Fateev and A.B. Zamolodchikov,
{\em Phys. Lett.} {\bf 87B} (1979)  389.
\bibitem{FZ} V. A. Fateev and A.B. Zamolodchikov, {\em Int. J. Mod. Phys.}
{\bf A5} (1990) 1025.
\bibitem{DDV} C. Destri and H.J. De Vega, {\em Phys. Lett.} {\bf B233} (1989)
336.
\bibitem{FKM} P.G.O. Freund, T.R. Klassen and E. Melzer, {\em Phys. Lett.}
{\bf B229} (1989) 243.
\bibitem{PD}P.E. Dorey, {\em Nucl. Phys.} {\bf B358} (1991) 654;
 {\em Nucl. Phys.} {\bf B374}
(1992) 741.
\bibitem{Musrep} G. Mussardo,  {\em Phys. Reports} {\bf 218} (1992) 215.
\bibitem{FJKO} A. Fring, P.R.  Johnson , M.A.C. Kneipp and D.I. Olive, \NP
{\bf B430} [FS] (1994) 597.
\bibitem{OTU}   D.I. Olive, N. Turok and J. Underwood, {\em Nucl. Phys.} {\bf
B409}  (1993) 509.
\bibitem{KM} T.R. Klassen and E. Melzer, \NP {\bf B350} (1991) 635.
\bibitem{Nie} M. Niedermeier, \NP {\bf B424} (1994) 184.
\bibitem{JackiwWoo} R. Jackiw and G. Woo, {\em Phys. Rev.} {\bf D12} (1975)
1643.
\bibitem{BFOP} J.A. Swieca, {\em Fortschr. der Physik} {\bf 25} (1977) 303;
B. Schroer and J.A. Swieca, \NP {\bf B121} (1977) 505;
M. Sato, T. Miwa and M. Jimbo, {\em Proc. Jap. Acad.} {\bf 53A} (1977)
6,147,153.
\bibitem{KT}
R. K\"oberle, V. Kurak and J.A. Swieca, {\em Phys Rev.} {\bf D20} (1979) 897;
M. Karowski and H.J.Thun, \NP {\bf B190} (1981) 61.
\bibitem{DFJ} V.G. Drinfel'd, {\em Soviet Math. Dokl.} {\bf 36} (1988) 212;
I.B. Frenkel and N. Jing, {\em Proc. Natl. Acad. Sci. USA} {\bf 85} (1988)
9373.
\bibitem{Mansfield} P. Mansfield, {\em Nucl. Phys.} {\bf B222} (1983), 419.
\bibitem{Sug}  H. Sugawara, {\em Phys. Rev.} {\bf 170} (1968) 1659.
\bibitem{Somm}  C.M. Sommerfield, {\em Phys. Rev.} {\bf 176} (1968) 2019.

\end{thebibliography}
\end{document}